\documentclass[10pt,aps,prb,superscriptaddress,
twocolumn,showpacs,floatfix,notitlepage]{revtex4-2}
\usepackage{amsmath}
\usepackage{amsfonts}
\usepackage{amssymb}
\usepackage{hyperref}
\usepackage{graphicx}
\usepackage{color}
\usepackage{xcolor}
\usepackage[mathscr]{euscript}
\usepackage{natbib}
\usepackage{rotating}
\usepackage[utf8]{inputenc}
\usepackage[T1]{fontenc}
\usepackage{soul}

\hypersetup{colorlinks=true,linkcolor=magenta
  ,citecolor=magenta, urlcolor=magenta}

\begin{document} 
\flushbottom 
\title{ Non-Hermitian  synthetic lattices with light-matter coupling}

\author{Amir Rahmani}
\affiliation{%
	Institute of Physics Polish Academy of Sciences, Al. Lotnik\'{o}w 32/46, 02-668 Warsaw, Poland}%

\author{Mateusz K\k{e}dziora}
\affiliation{Institute of Experimental Physics, Faculty of Physics, \\University of Warsaw, ul. Pasteura 5, PL-02-093 Warsaw, Poland}

\author{Andrzej Opala}
\affiliation{%
	Institute of Physics Polish Academy of Sciences, Al. Lotnik\'{o}w 32/46, 02-668 Warsaw, Poland}%
 \affiliation{Institute of Experimental Physics, Faculty of Physics, \\University of Warsaw, ul. Pasteura 5, PL-02-093 Warsaw, Poland}

\author{Michał Matuszewski}
\affiliation{%
	Institute of Physics Polish Academy of Sciences, Al. Lotnik\'{o}w 32/46, 02-668 Warsaw, Poland}%
\date{\today}

\begin{abstract} We propose that light-matter coupling can be used to realize synthetic lattices. In particular, we consider a one-dimensional chain of  exciton-photon sites to create a comb lattice that exhibits a transition from a  flat band to a finite mass dispersion by tuning site-dependent light-matter coupling. Moreover, in a non-Hermitian system with gain and loss, the flat band phase is much more robust and the transition is accompanied by the appearance of exceptional points in the complex energy spectrum.  We demonstrate that by engineering the light-matter coupling in the synthetic lattice, one can explore various phases in the lasing regime. Our proposal paves the way for studying non-Hermitian systems in higher dimensions.
  
\end{abstract}

\maketitle

\section{Introduction}
Lattice models are ubiquitous in physics, with applications ranging from approximations of real physical systems to efficient tools in  theoretical research such as lattice gauge theory. In the context of quantum simulation, synthetic  lattices are  highly beneficial as a method for 
using the internal degree of freedoms~\cite{Boada12} or external states~\cite{price17} to investigate nontrivial topology and higher dimensions
~\cite{Boada12,ozawa2019topological}. The formation of different lattice configurations may be implemented by using spin\cite{Celi2014}, Rydberg states\cite{Kanungo2022}, orbital angular momentum states\cite{Floss2015, Sundar2018} and photonic frequency comb~\cite{Lin16}.


Light-matter coupling may provide an advantage  in investigations of non-Hermitian physics~\cite{Yuto2020} due to flows of energy or particles to and from a system. A key feature of non-Hermitian systems is the presence of complex eigenvalues, which can lead to a variety of interesting phenomena such as non-reciprocal transport~\cite{Mandal2022}, topological phases~\cite{Gong18}, and exceptional points~\cite{Miri2019}. A possible non-Hermitian system is a lattice with gain and loss at its sites~\cite{szameti11,Takata18}.  Here we explore how light-matter coupling can provide an additional degree of freedom for creating a synthetic lattice. We study a simple model of a one-dimensional lattice in which the light-matter coupling can be manipulated. As such, we provide an example of a comb lattice. In one dimension, the lattice can be obtained from a two-leg ladder lattice~\cite{Rui22} by removing one leg  while keeping all other connections and couplings unchanged. Such lattices have been explored in the context of random walks~\cite{Pierre2016} and modeling chemical compounds~\cite{Shida07}.

 Light can couple to matter both weakly or strongly, in the latter case leading to hybrid light-matter quasiparticles called polaritons~\cite{Hopfield1958,Weisbuch1992}. 
Lattices of photonic nodes can be created by structuring the sample itself by deposition of a patterned layer on top of a cavity, sample etching \cite{Jacqmin2014} or etch-and-overgrowth\cite{Harder2021} procedures, as well as using a spatial light modulator to excite selected nodes\cite{Pickup2020}. The ease of creating arbitrary geometries and potential landscapes in polariton systems opened the way for studying a variety of Hamiltonian models, including Lieb lattices \cite{Whittaker2018,Baboux2016,Li2018,sun2018excitation,whittaker2021optical,baboux2018unstable,Fontaine2022,scafirimuto2021tunable,Klembt_Flatland,deuar2021fully}, Kagome\cite{Harder2021}, and honeycomb \cite{Jacqmin2014,Milicevic2015} lattices, as well as one-dimensional \cite{st2017lasing,Pieczarka:21} and two-dimensional~\cite{Klembt2018} topological systems. 
Owning to the possibility to manipulate gain and loss~\cite{fraser2022independent,hu2022grating} polaritons are an ideal system for studying non-Hermitian effects~\cite{Mandal22}.  A growing interest in such systems is directly related to the possibility of exploring the physics of exceptional points \cite{Richter2019,Miri2019, Su2021,Comaron20}, parity-time symmetry \cite{Ozdemir2019} or skin effect~\cite{Xueyi2020,Xu2021, Mandal2022,XU21}.

In this work, we propose that  light-matter coupling in the intermediate regime between strong and weak coupling can be exploited to introduce a synthetic lattice with nontrivial properties. The considered lattice, schematically shown in Fig.~\ref{fig:1}, is synthetic in the sense that it can be engineered to possess specific features such as lasing and flat bands. The fact that photon and exciton modes in the same micropillar couple allows to describe the system formally as a one-legged comb model. This allows investigation of the physics of a Lieb (Stub) lattice, including the appearance of a flat band. We show that site-dependent tuning of light-matter coupling strength allows the observation of the transition from a flat band to a dispersive spectrum.

Moreover, by analyzing the properties of the system in the more realistic non-Hermitian regime with gain and loss included, we find that the difference in loss rates between excitons and photons leads to a great enhancement of robustness of the flat band phase, and the appearance of a new flat band in a strongly non-Hermitian case. These effects are shown to persist in the nonlinear regime of lasing, where the signatures of a flat band and associated compact localized states~\cite{Leykam2018} (CLS) can be observed in a state resulting from a long-time evolution starting from a random initial condition.

Our results show that manipulating the light-matter coupling strength in lattices opens the way for the experimental realization of a new kind of synthetic  lattices that allow to explore strongly non-Hermitian physics. Our model can be easily extended to higher dimensions or a larger number of light and matter states, such as orthogonal light polarizations or higher exciton  states~\cite{piketka20172,piketka2017doubly}. It can be implemented in other physical platforms such as cavity quantum electrodynamics~\cite{mirhosseini2019cavity} and coupled atom-light systems~\cite{muniz2020exploring}.  We believe that our proposal may lead to engineering highly dimensional extended lattices, providing an ideal platform for simulators of complex non-Hermitian models.


\section{Model}\label{sec:model} 
\begin{figure}[h!]
  \centering
  \includegraphics[width=\linewidth]{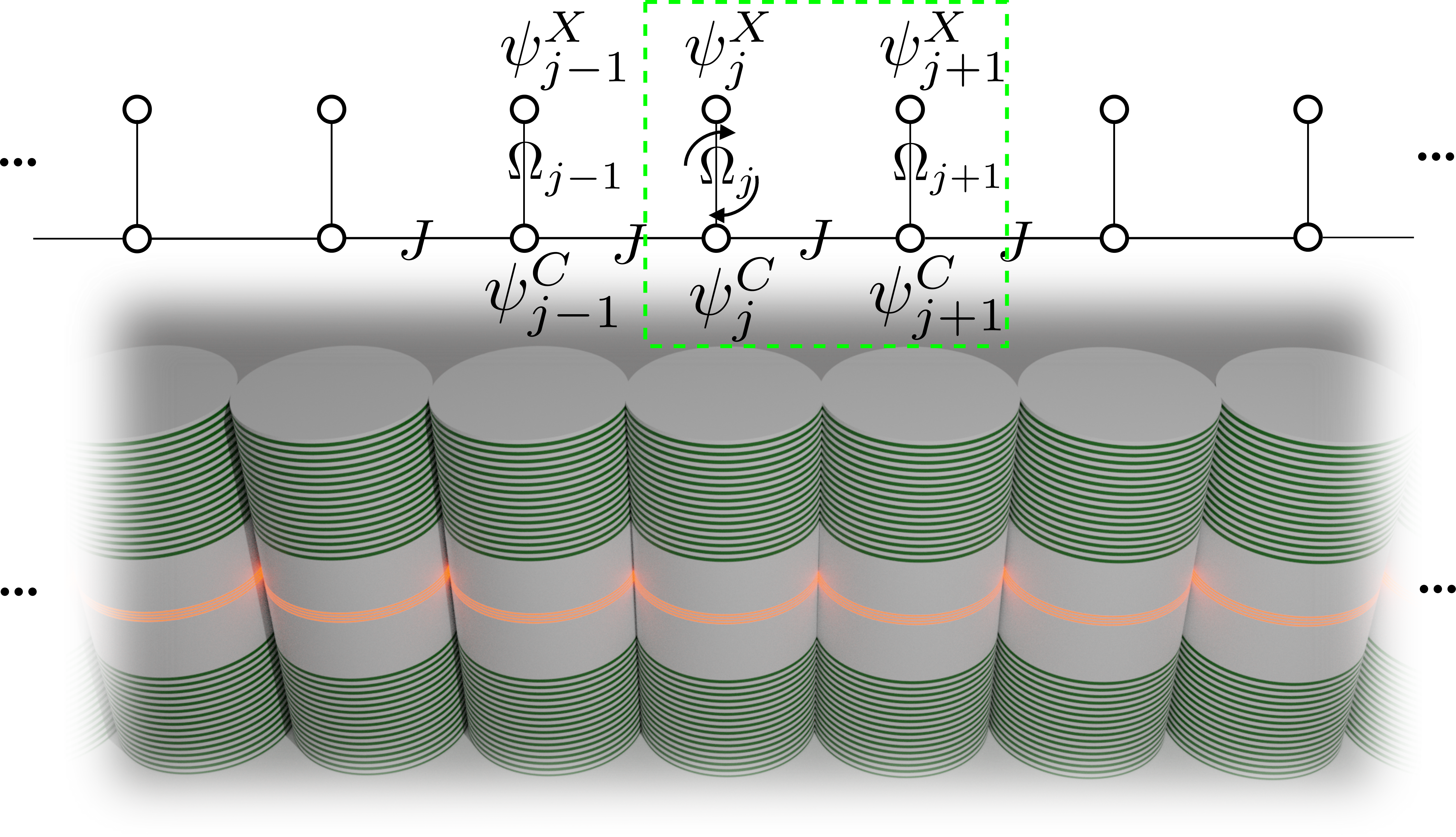}
  \caption{Scheme of our model. (Top) At lattice site $j$ photonic ($\psi_j^C$) and excitonic ($\psi_j^X$) states are  coupled. The neighboring photonic sites are coupled to each other with coupling rate $J$. A possible choice of a unit cell is marked by the green square. (Bottom) In practice the model can be implemented in a lattice of coupled micropillars. }
  \label{fig:1}
\end{figure}

The lattice described by our model shown in Fig.~\ref{fig:1}, consists of $2N$ micropillars ($N$ is the cell index, with two micropillars per unit cell) forming a one-dimensional chain with lattice constant $a$. Each site is described with two components, where we label exciton components with superscript $X$ and photonic components with superscript $C$.
We assume that  there is finite and uniform coupling between the sites via photonic components, while exciton components are not coupled with each other due to the much larger exciton effective mass. Photon and exciton modes in the same site are coupled with each other with the strength corresponding to Rabi frequencies which vary from site to site.  We consider a staggered distribution of couplings denoted by $\Omega_1$ in odd-numbered sites and $\Omega_2$ in even-numbered sites, where $\Omega_2 \leq\Omega_1$. In particular, such a position-dependent light-matter coupling can be realized in practice using site-dependent external pumping, which induces saturation of Rabi coupling~\cite{Ciuti_saturation}. Photon tunneling between photon modes localized in neighboring micropillars is described with $J$. As result, the model describing the system can be considered effectively two-dimensional with an additional two-site exciton-photon degree of freedom.

\label{sec:dwc983hf348fhuefe}
Our model equations can be presented in terms of four fields inside a unit cell. Denoting each cell with index $n=1,2,3,\cdots N$, we have $2N$ micropillars, for which we introduce fields $\psi^{C}_{j},\psi^{X}_{j}$ corresponding to photon and exciton components at site $j$, where $j=1,2,\dots,2N$. The mean-field evolution equations are
\begin{subequations}\label{eq:jdh98ey3hwedhw8dfded}
\begin{align}
    i\partial_t\psi^{C}_{j}=&-i \gamma_C\psi^{C}_{j}+J(\psi^{C}_{j-1}+\psi^{C}_{j+1})+\Omega_j\psi^{X}_{j}\,,\\
    i\partial_t \psi^{X}_{j}=&-i\gamma_X\psi^{X}_{j}+g\hbar^{-1}|\psi^{X}_{j}|^2\psi^{X}_{j}+\Omega_j\psi^{C}_{j}\,,
\end{align}
\end{subequations}
where $\Omega_j=\Omega_1$ for odd $j$ and  $\Omega_j=\Omega_2$ for even $j$, $\gamma_C (\gamma_X)$ denotes the decay rate from the photon (exciton) modes, and  $g$ is the nonlinear coefficient. Here $g$ is a complex number taking into account
both exciton-exciton interaction in its real part and gain saturation effect in its imaginary part~\cite{Comaron20}. In the case of periodic boundary conditions we assume $\psi_{N+1}^C=\psi_1^C$. For simplicity, we consider the case where there is no exciton-photon detuning at any site. 


To calculate the spectrum of the system, we consider the eigenvalue equation  $\Tilde{H}\Tilde{\Psi}=E(k)\tilde{\Psi}$ at $g=0$, where
$\tilde{\Psi}$ is a plane wave solution with momentum $k$.
We obtain
\begin{align}\label{eq:kjdwh9843hfu3eh}
   \tilde{H}= \begin{pmatrix}
-i\hbar\gamma_C & \hbar\Omega_1 &\hbar J(1+e^{-ika}) & 0\\
\hbar\Omega_1&-i\hbar\gamma_X&0&0\\
\hbar J(1+e^{ika}) & 0 & -i\hbar\gamma_C&\hbar\Omega_2\\0&0&\hbar\Omega_2&-i\hbar\gamma_X
\end{pmatrix}\,.
\end{align}
In the following, we will analyze in detail the properties of the system in both linear and nonlinear regimes. 

\begin{figure}[hbt]
  \centering
  \includegraphics[width=\linewidth]{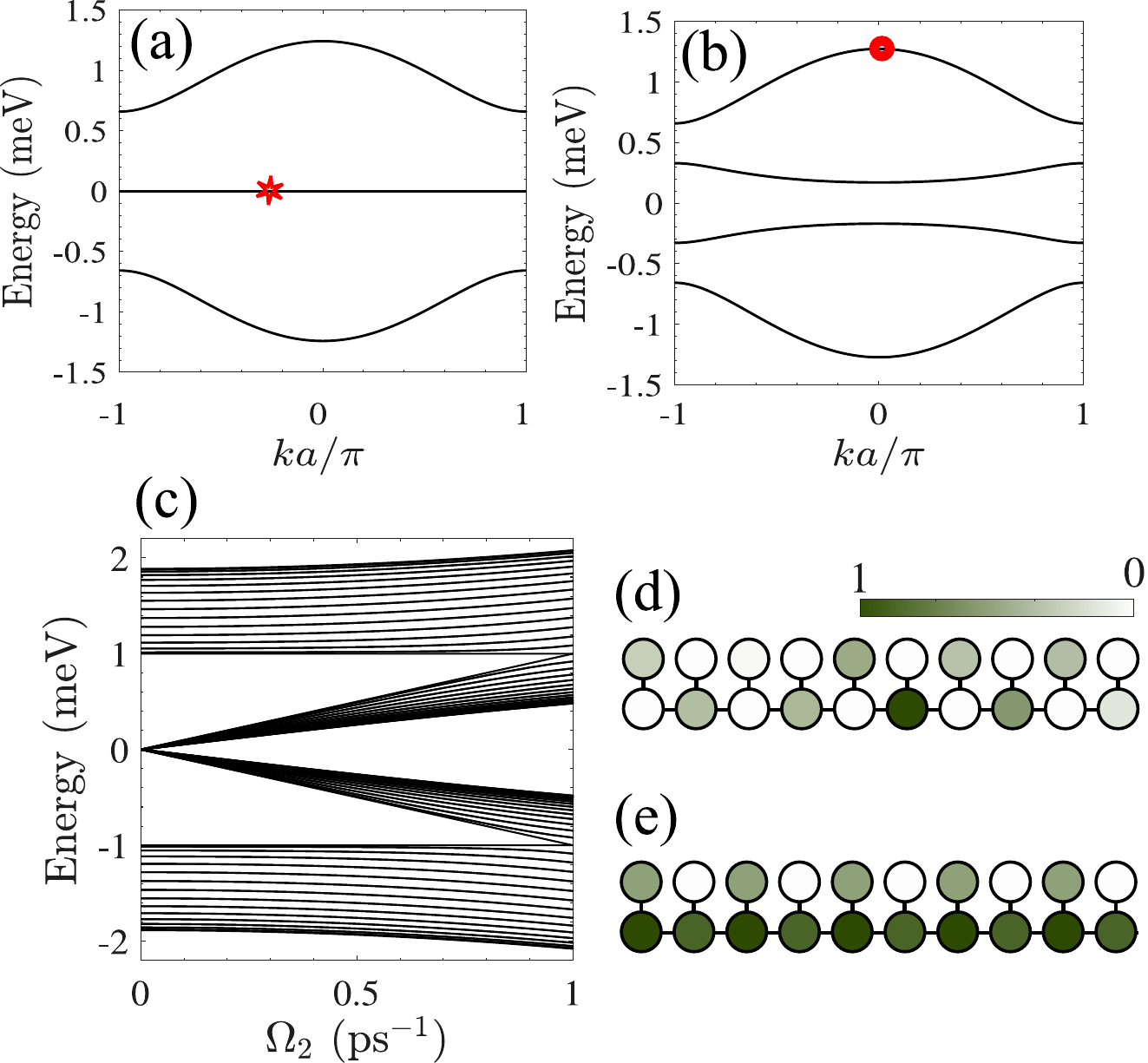}
  \caption{Examples of spectra and eigenstates in the linear Hermitian case. (a)-(b) Energy spectra in momentum space for an infinite system. In (a) we assume $\Omega_2=0$ (Stub lattice). In this case there are two degenerate flat bands with $E=0$, one corresponding to the Lieb lattice flat band, and another corresponding to isolated exciton sites $\psi^X_{2n}$. In (b) we have $\Omega_2=0.4~\mathrm{ps}^{-1}$, and the flat bands split into two dispersive bands. Panel (c) shows  eigenenergy spectrum in function of $\Omega_2$ in a finite system with $N=30$. (d),(e) Examples of density distributions for eigenstates corresponding to the flat band in (a) and dispersive band in (b), respectively. Corresponding points are marked with a red star and 
a circle in (a) and (b). Color scale is in arbitrary units. Other parameters are  $\Omega_1=1~\mathrm{ps}^{-1}$, $J=0.8~\mathrm{ps}^{-1}$. }
  \label{fig:fig2}
\end{figure}

\section{Hermitian case}\label{SEC:JDSFH832QYIQWEDW9E}
For the sake of clarity, we start our study with the linear Hermitian case $\gamma_C=\gamma_X=0$ and $g=0$. When $\Omega_2=0$, even exciton nodes $\psi^X_{2n}$ are completely isolated from all other nodes. The remaining photonic nodes and odd exciton nodes form the so-called one-dimensional Lieb lattice (also called Stub lattice) model~\cite{Baboux2016,Casteels2016,Real2017}. In this case one can find the dispersion equation analytically, $E(k)=0,\pm\hbar\sqrt{\Omega_1^2+2J^2(1+\cos(ka))}$. The model exhibits a flat band with $E=0$ and infinite mass separated by gaps from two dispersive bands, see Fig.~\ref{fig:fig2}(a).
Flat bands possess a number of intriguing physical phenomena, including compact localized states, sensitivity to perturbations and disorder, strongly correlated phases, and topological states~\cite{Leykam2018}. Note that in addition to the Lieb lattice flat band, the full model (\ref{eq:jdh98ey3hwedhw8dfded}) has a trivial flat band that corresponds to isolated exciton sites.

When the $\Omega_2$ parameter is nonzero, the two degenerate flat bands at $E=0$ split into two dispersive bands, as shown in Fig.~\ref{fig:fig2}(b) and Fig.~\ref{fig:fig2}(c). Hence,  in the Hermitian case, flat bands are present only in the limit of $\Omega_2=0$. On the other hand, in the symmetric case, $\Omega_1=\Omega_2$, the two bands with positive energy and the two bands with negative energy coalesce with each other, marking a transition to the uniform one-legged ladder model. In the intermediate regime $0<\Omega_2<1$, the two middle bands that emerged from the flat bands preserve some properties of the flat band eigenstates. In particular, the structure of the eigenstates in the middle bands resembles the CLS of the Lieb lattice, as in Fig.~\ref{fig:fig2}(d). In contrast, the eigenstates of the dispersive top and bottom bands resemble standard bulk states, see  Fig.~\ref{fig:fig2}(e).


%
\begin{figure}
  \centering
  \includegraphics[width=\linewidth]{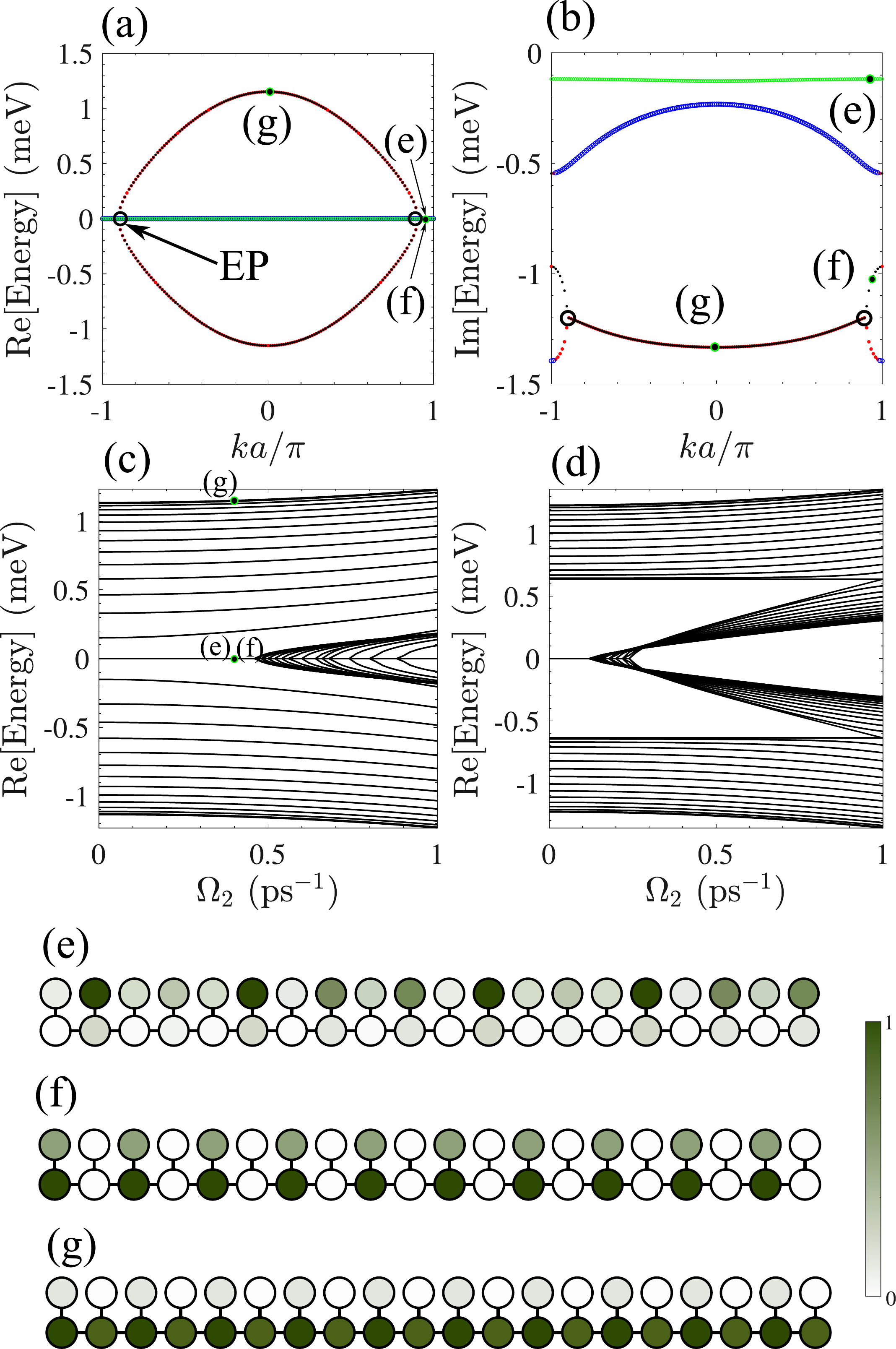}
  \caption{Linear non-Hermitian case. In (a) and (b) we show the real and imaginary parts of eigenvalues in function of quasimomentum for the same parameters as in Fig.~\ref{fig:fig2}(b), but with nonzero decay terms $\gamma_C=2.2~\mathrm{ps}^{-1}$ and $\gamma_X=0.1~\mathrm{ps}^{-1}$. In contrast to the Hermitian case, flat bands are present in the system for a non-zero coupling $\Omega_2=0.4~\mathrm{ps}^{-1}$. Exceptional points (EPs) are marked with a black open circle. In (c) and (d) we show real parts of eigenvalues in function of $\Omega_2$ for the fixed $\Omega_1=1~\mathrm{ps}^{-1}$. Qualitatively different spectra are shown in (d) for a smaller decay rate $\gamma_C=0.54~\mathrm{ps}^{-1}$. The eigenstates corresponding to the points marked in panels (a) and (b) are shown in panels (e)-(g).} 
  \label{fig:fig3}
\end{figure}

\section{Linear non-Hermitian case}\label{eq:hd8eyhiwendwedw8i}

We analyze the effect of nonzero decay $\gamma_C,$ $\gamma_X$ on the spectra and eigenstates of the system, keeping the interactions $g=0$. The results are presented in Fig.~\ref{fig:fig3}. It should be noted that adding a uniform decay rate $\gamma_C=\gamma_X>0$ would lead only to an addition of a constant imaginary part to the eigenvalues of the system, with no effect on the real part of the spectrum or the eigenfunctions. However, in real systems the decay rate of photons $\gamma_C$ is typically much higher than the decay rate for excitons $\gamma_X$. We find that this leads to a dramatic change in the spectra. An Example is shown in Figs.~\ref{fig:fig3}(a) and~\ref{fig:fig3}(b), where real and imaginary parts of eigenenergies are shown for the same parameters as in the Hermitian case of Fig.~\ref{fig:fig2}(b), but with $\gamma_C=2.2~\mathrm{ps}^{-1}$ and $\gamma_X=0.1~\mathrm{ps}^{-1}$. In contrast to the hermitian case, we find that despite nonzero $\Omega_2$, the system spectrum contains two flat bands in the entire Brillouin zone, while the dispersive bands develop small regions of flat dispersion at the two ends of the Brillouin zone. The transition from dispersive dependence at low quasimomentum to purely imaginary dependence at high quasimomenta is marked by the occurrence of exceptional points. We note that previously a purely imaginary dispersion was predicted to occur in the Bogoliubov excitation spectrum of a polariton condensate coupled to a reservoir~\cite{Wouters2018}. However, it occurrs at low momenta while the mini-flat bands  that appear in our model are present at high quasimomenta and the model does not assume condensation or include interactions with an uncondensed reservoir. 

The real part of energy eigenvalues is presented as a function of $\Omega_2$ in Figs.~\ref{fig:fig3}(c) and~\ref{fig:fig3}(d) for two values of the photon decay rate $\gamma_C$.  Upon changing the decay rate, exceptional points may emerge in the spectrum. While in the case of a high decay rate there are no energy gaps in the real part of the energy, in the low decay rate case the gaps are open at low values of $\Omega_2$ and the system spectrum is more similar to the Hermitian case of Fig.~\ref{fig:fig2}(c). Nevertheless, even in this case flat bands survive to a nonzero value of $\Omega_2$, in contrast to the Hermitian limit. This shows that including dissipation to the model, which occurs naturally in photonic systems, makes flat bands much more robust. Moreover, increasing $\gamma_C$ may induce exceptional points. At low decay rates, we have three distinct bands (including a flat band). This is shown in panel (d) for low values of $\Omega_2$. But by increasing the decay rate, the band gaps start decreasing, while exceptional points emerge at the gap closing. An example of a gapless spectrum is shown in panel (c). In general, increasing decay rate leads to coalescence of eigenvalues which gives rise to exceptional points and the disappearance of the gap. At larger values of the decay rate, exceptional points are present for any value of $\Omega_2$, and so the real part has no gap in panel (c). However, for a lower value of the decay rate the exceptional points may appear only at certain range of $\Omega_2$. For this reason, the spectrum in panel (d) displays band gaps.

Finally, in Figs.~\ref{fig:fig3}(e)-(g) we show three examples of density distributions of eigenstates corresponding to different bands. The eigenstate of the flat band shown in Fig.~\ref{fig:fig3}(e) preserves the characteristic  structure of the CLS of the Hermitian model, with all odd photonic sites being empty. In contrast, the state from the flat mini-band at high quasimomentum in Fig.~\ref{fig:fig3}(f) is characterized by a complementary pattern, with all even photonic sites being empty. An eigenstate from the dispersive band in the center of the Brillouin zone depicted in Fig.~\ref{fig:fig3}(g) shows no particular structure of photonic states, which are distributed more or less uniformly, with only a small admixture of excitonic states.
\begin{figure}[h]
            \centering
  \includegraphics[width=\linewidth]{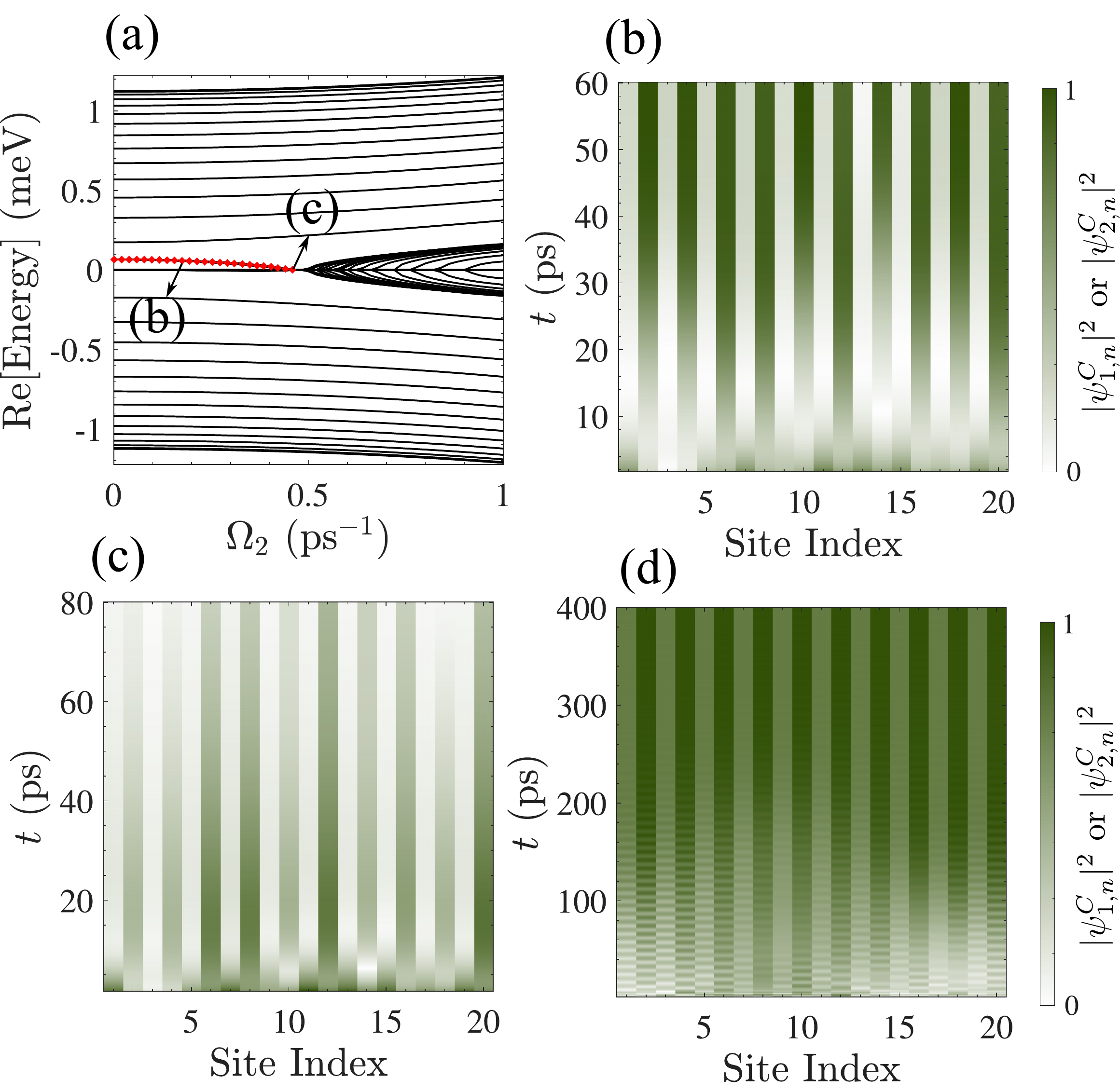}
 \caption{ (a) Real part of energy of excitations in a nonlinear steady state in function of $\Omega_2$  for $N=30$. The system is in the weakly nonlinear regime. Parameters are  $\gamma_C=2.2~\mathrm{ps}^{-1}$ and  $\gamma_X=-0.1~\mathrm{ps}^{-1}$, other parameters as in Fig.~\ref{fig:fig3}. Red dots show the energy of the linear excitation mode with zero imaginary part. All other excitations have energies with negative imaginary parts. In panels (b)-(d) we show the evolution of densities in photonic sites. In panel (b) we used $\Omega_2=0.18~\mathrm{ps}^{-1}$, which corresponds to a non-zero real eigenenergy, see panel (a). In panel (c) we used $\Omega_2=0.46~\mathrm{ps}^{-1}$, for which the real part of the energy is very near the zero-energy flat band. The resulting density shows signatures of random localization. Panel (d) shows an example of dynamics in a weakly dissipative case where we used $\gamma_C=0.54~\mathrm{ps}^{-1}$,   $\gamma_X=-0.12~\mathrm{ps}^{-1}$, and $\Omega_2=0.85~\mathrm{ps}^{-1}$. Here the system reaches the steady state after initial Rabi oscillations. We assume the nonlinear parameter $\hbar g=0.005-0.005i~$$\mathrm{meV^2~ps}$ which consists of both density-preserving interactions and density-dependent decay.}
  \label{fig:fig4}
        \end{figure}    
\section{Lasing}\label{se:jdf98etfw9e8df8we}
In the previous section we considered a lattice in the linear non-Hermitian regime. A natural extension is a system in the presence of nonlinearity, $g\neq 0$. Lasing corresponds to a steady state in the case of positive gain. To achieve the steady state regime,  we add a new feature to our dissipative model, that is, we consider a balance between gain and nonlinear decay. In this case particles escaping the lattice can be replenished by a pumping term. 

This can be analyzed simply by assuming $\gamma_X<0$ (as the pump source, eg. a nonresonant optical pump creating excitons) while $\gamma_C>0$ (photon decay). In this regime, we need to assume a nonzero-complex nonlinear term ($g\neq0$) in Eqs.~(\ref{eq:jdh98ey3hwedhw8dfded}) to reach a stable steady state. We  keep the associated parameters in the so called weak nonlinear regime when $|g||\psi_{1,n}|^2$ and $|g||\psi_{2,n}|^2$ are at least one order of magnitude lower than $\hbar \gamma_C$. As an aside, one possible source of disorder that may become important in our modeling is due to randomness in pumping, which is not considered here, but the results remain unchanged with a minor percent of disorder strength~\cite{subhaskar23}. We provide examples of the dynamics in the steady-state regime in Figure \ref{fig:fig4}.

 Additionally, we calculate the spectrum of small Bogoliubov excitations around the steady state. As we are dealing with a homogeneous case, in the steady state case we may consider a stationary solution in the form $\psi_{j}^{SC,SX}e^{-iE t/\hbar}$. The phase factor holds information about the eigenvalues of the model in real space. Indeed, by substituting the above ansatz in Eq.~(\ref{eq:jdh98ey3hwedhw8dfded}), one can find a set of time-independent equations
\begin{subequations}\label{eq:jdh9asd3hwedhw8dfded}
\begin{align}
    0=&(-i \hbar \gamma_C-E)\psi^{SC}_{j}+\hbar J(\psi^{SC}_{j-1}+\psi^{SC}_{j+1})+\hbar\Omega_j\psi^{SX}_{j}\,,\\
    0=&(-i\hbar \gamma_X-E)\psi^{SX}_{j}+\hbar\Omega_ j\psi^{SC}_{j}+g|\psi^{SX}_{j}|^2\psi^{SX}_{j}\,,
\end{align}
\end{subequations}
which can be solved to find the eigenvalues $E$. To this end, we solve model equations (Eq.~(\ref{eq:jdh98ey3hwedhw8dfded})) numerically by employing Runge-Kutta method, and use the corresponding excitonic occupations in the steady state solution $|\psi^{SX}|^2$ in the above equations.

In the steady state, we are interested in excitations for which the imaginary part of the energy eigenvalue $\mathrm{Im}[E]$ is zero, since for $\mathrm{Im}[E]>0$ the steady state is unstable, and for $\mathrm{Im}[E]<0$ the excitations decay in time. Excitation modes with $\mathrm{Im}[E]=0$ can be considered as discrete analogs of Goldstone modes. We assume that the parameter that can be manipulated externally is $\Omega_j$. As such, it would be useful to consider the variations of energies with $\Omega_2$ in a steady state.

Real parts of excitation eigenvalues are shown in Fig.~\ref{fig:fig4}(a). Here we assume that the interactions in a chain of micropillars are effectively  repulsive via the effect of 
exciton-exciton interactions. Considering the case when particle conserving interactions and gain saturation are of the same order of magnitude, we choose $\hbar g=0.005-0.005i$~$\mathrm{meV^2~ps}$ as our effective nonlinear parameter taking into account both conservative and dissipative nonlinear processes. We note that in comparison to the linear case in Fig.~\ref{fig:fig3}(c), at small $\Omega_2$ a new split band is present in Fig.~\ref{fig:fig4}(a). This splitting from the zero energy band is a nonlinear effect, that is, the nonlinear term  induces a positive onsite potential for the excitons, proportional to $\mathrm{Re}[g]|\psi^X_{j}|^2$. We note that for small $\Omega_2$ exciton sites which are indexed by $j=2n$ are almost isolated from the rest of the lattice. Since excitons at these sites do not experience photonic decay $\gamma_C$, their decay is controlled by the imaginary part of the nonlinear term $g|\psi^X_{j}|^2$. This term is much smaller than the decay of photonic states $\gamma_C$ in the case of the weakly nonlinear regime that we consider here. Hence, the $j=2n$ excitonic sites experience the highest gain, which results in a steady state where density is mostly localized at these sites at small $\Omega_2$. We show an example of steady state formation starting from a random initial condition in Fig.~\ref{fig:fig4}(b).

As $\Omega_2$ increases, these disconnected excitonic states ($\psi^X_{2n}$) couple strongly with the other states. The populations in the other sites in the steady state increase while the population in $\psi^X_{2,n}$ decreases. At some critical $\Omega_2$ the split band gap is closed and the real part  of the energy eigenvalue becomes zero, as visible in Fig.~\ref{fig:fig4}(a). At this point, the eigenstate of the system with the highest imaginary part of the energy belongs to $E=0$ flat band. This has a profound effect on the dynamics of the system. As shown in Fig.~\ref{fig:fig4}(c), states that are spontaneously formed from a random initial condition result in a different random density distribution after some certain time. This random density distribution is long lived and locally resembles the density of a flat band state. This can be seen as a reminiscence of flat band CLS of the linear Lieb model, see Fig.~\ref{fig:fig2}(d). Further increasing $\Omega_2$ with  other parameters fixed leads to a situation where no steady state exists since all the eigenstates have negative imaginary part. In this case any initial distribution decays to zero. A qualitatively different situation occurs where the photon decay is decreased, as shown in Fig.~\ref{fig:fig4}(d). In this case the system converges to a steady state with a periodic, regular density distribution with all sites occupied. The initial dynamics shows a clear signature of collective Rabi oscillations in the lattice, which decays after a sufficiently long time.
 
\section{conclusion}
 \label{sec:jsd89wehdiwhed}
In summary, we propose that a lattice system with coupled light and matter modes can be used to realize a synthetic  comb lattice by employing the internal degree of freedom due to light-matter coupling. This idea can be readily realized in exciton-polariton systems, where several methods of tuning light-matter coupling strength have been shown. We demonstrated that local engineering of light-matter coupling provides a way to explore a dissipative phase transition between the regimes of dispersive and flat band phases. The transition is accompanied by the appearance of exceptional points in the spectrum. Importantly, in the dissipative case the flat band regime is not restricted to a limiting case in the parameter space, but exists in a range of values of light-matter coupling, which makes the phenomenon much more robust than in the Hermitian case. We also showed that the existence of a flat band has a profound effect on the states of the system after long evolution in the regime of lasing, enabling a straightforward experimental observation. The proposed method can be generalized to higher dimensional lattices,  and the number of sites in the synthetic  lattice can be increased by considering additional light and matter states. This  opens the way to explore non-Hermitian systems in higher dimensions.
\section*{Acknowledgement}
AR, MK, and MM acknowledge support from National Science Center, Poland, Grant No. 2016/22/E/ST3/00045, A.O. acknowledges 2019/35/N/ST3/01379.
\section*{Appendix}
The dynamics of exciton-polaritons are known to be described by mean-field Gross-Pitaevskii equations:
\begin{subequations}
	\begin{align}
	i\hbar \partial_t \psi^C=&-i\hbar\gamma_C\psi^C+\hbar\Omega \psi^X\,,\label{adwoqyedqw86y}\\
	i\hbar \partial_t \psi^X=&(-\delta-i\hbar\Gamma_X)\psi^X+\hbar\Omega \psi^C+\alpha_1|\psi^X|^2\psi^X\nonumber\\&+g_R n_R\psi^X+i\hbar R n_R\psi^X\,,\label{adwoqasdyedqw86y}\\
	\partial_t n_R=&P-(\gamma_R+R|\psi^X|^2)n_R\label{adwoqyesdfdqw86y}\,,
	\end{align}
\end{subequations}
where $n_R$ is the density of the exciton reservoir, $P (\gamma_R)$ is the rate of pumping to (decay from) a reservoir. There are three parameters that take into account interactions: $\alpha_1$, $g_R$ and $R$. The energy detuning between the exciton and the photon field is given by $\delta$. By using the adiabatic approximation~\cite{PhysRevB.92.035311} one can assume that $n_R\approx \frac{P}{\gamma_R}(1-\frac{R|\psi^X|^2}{\gamma_R})$, the effective equation for the exciton field becomes
\begin{align}
i\hbar \partial_t \psi^X=&-i\hbar \gamma_X\psi^X+\hbar\Omega \psi^C+g|\psi^X|^2\psi^X\,,\label{ad22wdsoqasdyedqw86y}
\end{align}  
where we introduce $-i\hbar \gamma_X=-\delta-i\hbar\Gamma_X+P/\gamma_R(g_R+i\hbar R)$, $g=\alpha_1-\frac{g_R P R}{\gamma_R^2}-i\frac{\hbar R^2 P}
{\gamma_R^2}$. Equation (\ref{ad22wdsoqasdyedqw86y}) has been used to simulate the exciton field in the main text.
\bibliography{vc}
\end{document}